\documentclass[%
    reprint,
    twocolumn,
    nofootinbib,
    amsmath,
    amssymb,
    aps,
    prstab,
]{revtex4-2}

\usepackage[export]{adjustbox}
\usepackage{graphicx} 
\usepackage{dcolumn}
\usepackage{bm}
\usepackage[caption=false]{subfig}

\newcommand{\vect}[1]{\boldsymbol{\mathbf{#1}}}

\begin{document}

\title{High-dimensional maximum-entropy phase space tomography using normalizing flows}

\author{Austin Hoover}%
\email{hooveram@ornl.gov}
\affiliation{Oak Ridge National Laboratory, Oak Ridge, Tennessee 37830, USA}

\author{Jonathan C. Wong}%
\affiliation{Institute of Modern Physics, Chinese Academy of Sciences, Lanzhou 730000, China}

\date{\today}

\begin{abstract}

Particle accelerators generate charged particle beams with tailored distributions in six-dimensional position-momentum space (phase space). Knowledge of the phase space distribution enables model-based beam optimization and control. In the absence of direct measurements, the distribution must be tomographically reconstructed from its projections. In this paper, we highlight that such problems can be severely underdetermined and that entropy maximization is the most conservative solution strategy. We leverage \textit{normalizing flows}---invertible generative models---to extend maximum-entropy tomography to six-dimensional phase space and perform numerical experiments to validate the model's performance. Our numerical experiments demonstrate consistency with exact two-dimensional maximum-entropy solutions and the ability to fit complicated six-dimensional distributions to large measurement sets in reasonable time.
\end{abstract}

\maketitle

\section{Introduction}\label{sec:introduction}

\begin{figure}
    \centering
    \includegraphics[width=\columnwidth]{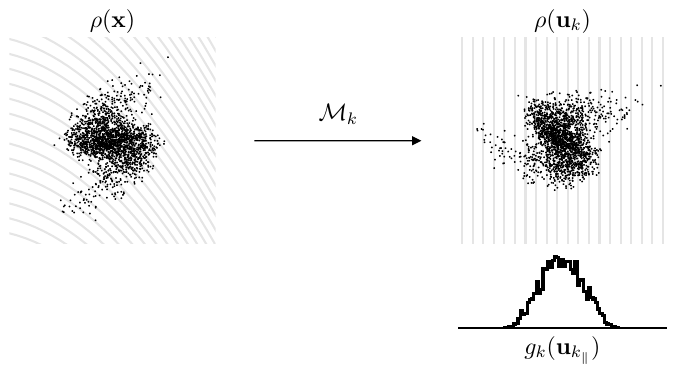}
    \caption{Generic phase space tomography setup. An initial phase space distribution $\rho(\vect{x})$ travels through an accelerator segment represented by the symplectic transformation $\vect{u}_k = \mathcal{M}_k(\vect{x})$ for measurement index $k$. Each projection $g_k(\vect{u}_{k_\parallel})$ of the transformed distribution is a different low-dimensional view of the initial distribution.}
    \label{fig:diagram}
\end{figure}

Particle accelerators generate charged particle beams with tailored distributions in position-momentum space (phase space). Measuring the phase space distribution in the accelerator enables model-based beam optimization and control and provides a valuable benchmark for simulation codes. In the absence of direct measurements \cite{Cathey_2018, Ruisard_2020, Hoover_2023}, the distribution must be reconstructed from its projections.\footnote{The 1D beam density can be measured by recording the secondary electron emission from a wire swept across the beam. Scintillating screens provide 2D projections of electron beams or low-intensity, low-energy hadron beams. 2D projections of higher energy hadron beams are only available from specialized diagnostics such as laser wires \cite{Liu_2020}.} Fig.~\ref{fig:diagram} illustrates a generic setup in which the beam is measured under varying accelerator conditions and reconstructed at a location upstream of the measurement device.

If the accelerator linearly transforms the phase space coordinates and does not couple the three planes of motion, one can reconstruct the 2D phase space distribution using conventional tomography algorithms. It is more challenging to reconstruct the 4D or 6D phase space distribution. Many conventional algorithms represent the distribution on a grid and face massive storage requirements as the phase space dimension scales \cite{Wolski_2020}. Several authors have developed new algorithms and diagnostics to sidestep this issue and fit 4D phase space distributions to 2D projections \cite{Hock_2013, Wang_2019, Wolski_2020, Marchetti_2021, Wolski_2023, Roussel_2023}. Recent work has also explored extensions to 5D and 6D phase space \cite{Scheinker_2023, Jaster_2024, Roussel_2024_6d}.

An additional challenge is that high-dimensional reconstructions may be ill-posed; since the measured dimension is fixed, the set of feasible distributions (those consistent with the measurements) may proliferate with the phase space dimension. It is usually infeasible to compensate by exponentially increasing the number of measurements, as one is typically limited to tens of views because of slow diagnostic devices and limited beam time. Additionally, it is not yet clear how to derive the information-maximizing set of high-dimensional phase space transformations under given measurement conditions---and in any case, accelerator constraints place many transformations out of reach. 

To select a single solution from the feasible set, our strategy is to define a prior probability distribution over the phase space coordinates and update the prior to a posterior by incorporating the information in the measurements. Our information comes in the form of constraints, and we perform the update by maximizing a convex functional subject to these constraints. Under basic self-consistency requirements, the functional must be the relative entropy \cite{Presse_2013, Skilling_1991, Jaynes_2012, Giffin_2008}. Entropy maximization ensures that the posterior does not deviate from the prior unless forced to by the data. This is a conservative strategy that eliminates all spurious features from the reconstructed distribution.

Entropy maximization is not always feasible, especially in high dimensions, because it entails a highly nonlinear constrained optimization. Although a reliable exact maximum-entropy algorithm exists for 2D tomography, its computational complexity scales exponentially with the phase space dimension, rendering its extension to 6D prohibitively expensive at this time. In this paper, we leverage \textit{normalizing flows}---invertible generative models---to find approximate 6D maximum-entropy solutions. Our approach is a straightforward extension of two previous studies. Loaiza-Ganem, Gao, and Cunningham \cite{Loaiza_2016} first proposed the use of normalizing flows for entropy maximization subject to statistical moment constraints; we incorporate projection constraints using the differentiable physics simulations and projected density estimation proposed by Roussel et al. \cite{Roussel_2023} in the Generative Phase Space Reconstruction (GPSR) framework. We refer to the resulting approach as MENT-Flow.

We begin by deriving the form of the $n$-dimensional maximum-entropy distribution subject to $m$-dimensional projection constraints, following the analysis in \cite{Wong_2022}. We then discuss the shortcomings of existing maximum-entropy tomography algorithms when $n = 6$ and describe the flow-based solution. Finally, we perform numerical experiments to validate the model's reliability in 2D settings and examine the effects of entropic regularization in 6D tomography.

\section{Maximum entropy tomography} \label{sec:ment}

Let $\rho_*(\vect{x})$ be a prior probability distribution over the phase space coordinates $\vect{x} \in \mathbb{R}^n$. We wish to update the prior to a posterior $\rho(\vect{x})$ by maximizing the relative entropy
\begin{equation}\label{eq:entropy}
    H[\rho(\vect{x}), \rho_*(\vect{x})] =
    -\int{
        \rho(\vect{x}) 
        \log{ 
            \left( \frac{\rho(\vect{x})}{\rho_*(\vect{x})} \right)
        } 
        d\vect{x}
    }
\end{equation}
while enforcing consistency with a set of $m$-dimensional projections. We will refer to this problem as an $n$:$m$ reconstruction. 

We assume the $k$th measurement occurs after a symplectic transformation $\mathcal{M}_k: \mathbb{R}^n \rightarrow \mathbb{R}^n$. By splitting the transformed coordinates 
\begin{equation} \label{eq:forward}
    \vect{u}_k = \mathcal{M}_k(\vect{x})
\end{equation}
into a projection axis $\vect{u}_{k_\parallel} \in \mathbb{R}^m$ and orthogonal integration axis $\vect{u}_{k_\perp} \in \mathbb{R}^{n - m}$, we can write the constraints as
\begin{equation} \label{eq:constraints}
    G_k \left[ \rho(\vect{x}) \right] 
    = g_k(\vect{u}_{k_\parallel}) - \tilde{g}_k(\vect{u}_{k_\parallel})
    = 0,
\end{equation}
where $g_k(\vect{u}_{k_\parallel})$ are the \textit{measured} projections and
\begin{equation} \label{eq:simulated-projections}
\begin{aligned}
    \tilde{g}_k(\vect{u}_{k_\parallel}) 
    =
    \int \rho ( \vect{x}(\vect{u}_k) ) d\vect{u}_{k_\perp}
\end{aligned}
\end{equation}
are the \textit{simulated} projections. The form of the maximum-entropy posterior distribution can be derived from a new functional
\begin{equation}
    \Psi
    =
    H[\rho(\vect{x}), \rho_*(\vect{x})]
    + 
    \sum_{k}^{} {
        \int
        \lambda_{k}(\vect{u}_{k_\parallel})
        G_k \left[ \rho(\vect{x}) \right]
        d\vect{u}_{k_\parallel}
    },
\end{equation}
where $\lambda_k(\vect{u}_{k_\parallel})$ are Lagrange multipliers \cite{Mottershead_1996}. Enforcing zero variation of $\Psi$ with respect to $\rho(\vect{x})$ and $ \lambda_k(\vect{u}_{k_\parallel})$ gives
\begin{equation} \label{eq:ment_solution}
\begin{aligned}
    \rho(\vect{x}) 
    &= 
    \rho_*(\vect{x}) 
    \prod_{k} \exp{ \left( \lambda_k(\vect{u}_{k_\parallel} (\vect{x})) \right) } 
    \\
    &= 
    \rho_*(\vect{x}) 
    \prod_{k} h_k ( \vect{u}_{k_\parallel} (\vect{x}) ).
\end{aligned}
\end{equation}
where we have defined $h_k(\vect{u}_k) = \exp(\lambda_k(\vect{u}_k))$. Substituting Eq.~\eqref{eq:ment_solution} into Eq.~\eqref{eq:constraints} generates a set of coupled nonlinear integral equations from which $h_k$ are to be solved.

\subsection{MENT}\label{subsec:MENT}

The MENT algorithm \cite{Minerbo_1979, Dusaussoy_1991, Mottershead_1996, Wong_2022} leverages a Gauss-Seidel relaxation method to optimize the Lagrange functions in Eq.~\eqref{eq:ment_solution}. After initializing the distribution to the prior within the measurement boundaries:
\begin{equation} \label{eq:ment_initialization}
    h_k(\vect{u}_{k_\parallel}) = 
    \begin{cases}
        1, & \text{if}\ g_k(\vect{u}_{k_\parallel}) > 0  \\
        0, & \text{otherwise}
    \end{cases},
\end{equation}
the Lagrange functions are updated as
\begin{equation} \label{eq:ment_gauss_seidel}
    h_k( \vect{u}_{k_\parallel} ) \leftarrow
    h_k( \vect{u}_{k_\parallel} )
    \left(
        1 + \omega
        \left(
            \frac
            {g_k(\vect{u}_{k_\parallel})}
            {\tilde{g}_k(\vect{u}_{k_\parallel})}
            - 1
        \right)
    \right)
\end{equation}
where
\begin{equation} \label{eq:ment_sim_projections}
    \tilde{g}_k(\vect{u}_{k_\parallel}) 
    =
    \int{
        \rho_*(\vect{x}(\vect{u}_k))
        \prod \limits_{j}
        h_j ( \vect{u}_{j_\parallel} (\vect{u}_k) )
        d\vect{u}_{k_\perp}
    }
\end{equation}
are the simulated projections and $0 < \omega \le 1$ is a learning rate \cite{Mottershead_1996}. The updates are performed in order ($k = 1, 2, 3, \dots$), and each updated $h_k$ is immediately used to simulate the next projection. One epoch is completed when all functions are updated. The iterations in Eq.~\eqref{eq:ment_gauss_seidel} converge \cite{Minerbo_1979, Dusaussoy_1991}. 

MENT maximizes entropy by design: fitting the data generates an exact solution to the constrained optimization problem. MENT is also efficient: it stores the exact number of parameters needed to define the maximum-entropy distribution and typically converges in a few epochs. Finally, MENT is essentially free of hyperparameters.

The MENT formulation above is valid for $n$:$m$ tomography, but the integrals in Eq.~\eqref{eq:ment_sim_projections} limit the value of $n$ in practice. Ongoing work aims to demonstrate efficient implementations when $n = 4$ \cite{Wong_2022, Tran_2023_ipac}. Extension to $n = 6$ may be possible, but it has yet to be demonstrated, and the runtime would likely be quite long if there were many high-resolution measurements. Even if the algorithm converged, sampling particles from the posterior (Eq.~\eqref{eq:ment_solution}) would be a nontrivial extra step.

\subsection{MENT-Flow} \label{subsec:ment-flow}

In the absence of a method to directly optimize the Lagrange functions in Eq.~\eqref{eq:ment_solution}, we may try to minimize the loss function
\begin{multline}\label{eq:loss}
    L = 
    - H[\rho(\vect{x}), \rho_*(\vect{x})]
    +
    \mu \sum_{k} {
        D[
            {g}_k(\vect{u}_{k_\parallel}), 
            \tilde{g}_k(\vect{u}_{k_\parallel})
        ]
    }
\end{multline}
for an increasing sequence of penalty parameters $\mu$. Here, $D[{g}_k(\vect{u}_{k_\parallel}), \tilde{g}_k(\vect{u}_{k_\parallel})]$ is a non-negative number quantifying the discrepancy between the measured and simulated projections, which we choose to be the Kullback-Leibler (KL) divergence. Exact solutions may require $\mu \rightarrow \infty$, but approximate solutions obtained with finite $\mu$ are often sufficient. 

The above approach requires us to represent the distribution using a finite set of parameters, $\vect{\theta}$. Grid-based representations become expensive when $n \geq 4$. An attractive alternative is to directly predict the value of $\rho(\vect{x})$ up to a normalization constant; however, computing the distribution's entropy (Eq.~\eqref{eq:entropy}) and projections (Eq.~\eqref{eq:simulated-projections}) would require expensive integration or Monte Carlo sampling. We might instead define the distribution indirectly via the transformation
\begin{equation} \label{eq:flow_forward}
    \vect{x} = \mathcal{F}(\vect{z}; \vect{\theta}),
\end{equation}
where $\mathcal{F}: \mathbb{R}^{n'} \rightarrow \mathbb{R}^{n}$ is a map parameterized by $\vect{\theta}$, and $\mathbf{z} \in \mathbb{R}^{n'}$ is a random variable drawn from a base distribution $\rho_0(\vect{z})$ defined in a ``normalized'' or ``latent'' space. The base distribution is typically a Gaussian. Sampling from $\rho(\vect{x})$ reduces to sampling from $\rho_0(\vect{z})$ and applying the unnormalizing transformation in Eq.~\eqref{eq:flow_forward}. Thus, Eq.~\eqref{eq:flow_forward} defines a \textit{generative model}. In most generative models, $\mathcal{F}$ is a neural network trained to learn an unknown distribution from data samples \cite{Bond_2021}.

Roussel et al. \cite{Roussel_2023} showed that generative models can also be trained to match \textit{projections} of the unknown distribution. To train the model via gradient descent, the transformations from the base distribution to the measurement locations must be differentiable:
\begin{equation}
    \vect{u}_k  = \mathcal{M}_k ( \mathcal{F}(\vect{z}; \vect{\theta}) ).
\end{equation}
This is possible using a differentiable beam physics simulation \cite{Kaiser_2024} to represent $\mathcal{M}_k$. The calculation of the projected density $\tilde{g}_k(\vect{u}_k)$ in Eq.~\eqref{eq:simulated-projections} must also be differentiable. This is possible using 1D or 2D kernel density estimation. It is, however, difficult to maximize the entropy without access to the density $\rho(\vect{x})$.\footnote{Roussel et al. \cite{Roussel_2023} proposed to maximize the \textit{emittance}, or root-mean-square (rms) volume, $\varepsilon = {|\vect{\Sigma}|}^{1/2}$, where $\vect{\Sigma} = \langle \vect{x}\vect{x}^T \rangle$ is the $n \times n$ covariance matrix of second order moments, as a proxy for the entropy. For certain distributions, the logarithm of the emittance is proportional to the entropy, but this is not true in general. Like entropy maximization, emittance maximization removes unnecessary linear correlations from the reconstructed distribution. However, it cannot remove nonlinear correlations, as the emittance depends only on second-order moments. Furthermore, the maximum-emittance distribution is not unique. In linear systems, the covariance matrix is typically overdetermined by the tomographic measurements, i.e., all distributions that fit the data have the same emittance. Particle-based entropy estimates based on k nearest neighbors \cite{Ao_2022_entropy} may perform better.}

A \textit{normalizing flow}, or simply \textit{flow}, follows the same paradigm but provides access to the probability density. A normalizing flow is a differentiable map $\mathcal{F}: \mathbb{R}^n \rightarrow \mathbb{R}^n$ with a differentiable inverse $\mathcal{F}^{-1}$ \cite{Papamakarios_2021}. These properties ensure we can compute the change in probability density under the transformation in Eq.~\eqref{eq:flow_forward}:
\begin{equation} \label{eq:flow_change_of_variables}
    \log{\rho(\vect{x})} = \log{\rho_0(\vect{z})} - \log{\left| {\det J_{\mathcal{F}}(\vect{z})} \right|},
\end{equation}
where
\begin{equation} \label{eq:flow_jacobian}
    J_{\mathcal{F}}(\vect{z}) = 
    \frac{d\mathcal{F}}{d\vect{z}} = 
    \begin{bmatrix}
        \frac{\partial{\mathcal{F}_1}}{\partial{z_1}} & \dots &  \frac{\partial{\mathcal{F}_1}}{\partial{z_n}} \\
        \vdots & \ddots & \vdots \\
        \frac{\partial{\mathcal{F}_n}}{\partial{z_1}} & \dots &  \frac{\partial{\mathcal{F}_n}}{\partial{z_n}} \\
    \end{bmatrix}.
\end{equation}
is the $n \times n$ Jacobian matrix of $\mathcal{F}$, accounting for volume change, and $\vect{z} = [z_1, \dots , z_n]^T$. To compute the probability density at $\vect{x}$, we flow backward and multiply the base distribution at $\vect{z}$ by the absolute value of the Jacobian matrix determinant. To generate samples $\{ \vect{x}_i \}$, we sample points $\{ \vect{z}_i \}$ from the base distribution and unnormalize them by flowing forward. We also obtain the probability density $ \{\rho(\vect{x}_i) \}$ at each sampled point by tracking the Jacobian matrix determinant during this forward pass.

The ability to generate particles \textit{and} evaluate the probability density at each particle is useful for computing expected values. Given $N$ samples $\{ \vect{x}_i \}$ from $\rho(\vect{x})$, the following expression is an unbiased estimate of the expected value of a functional $Q[\rho(\vect{x})]$:
\begin{equation} \label{eq:monte_carlo}
\begin{aligned}
    \mathbb{E} 
    \left[ Q[\rho(\vect{x})] \right]
    = \int \rho(\vect{x}) Q[\rho(\vect{x})] d\vect{x}
    \approx 
    \frac{1}{N} 
    \sum_{i=1}^{N}  Q[\rho(\vect{x}_i)].
\end{aligned}
\end{equation}
Since the entropy is the expected value of $\log ( \rho(\vect{x}) / \rho_*(\vect{x}))$, the following expression is an unbiased estimate of the entropy \cite{Loaiza_2016}:
\begin{equation} \label{eq:flow_entropy}
\begin{aligned}
    H[\rho(\vect{x}), \rho_*(\vect{x})]
    &\approx
    -\frac{1}{N} 
    \sum_{i=1}^{N} 
    \log ( \rho(\vect{x}_i) / \rho_*(\vect{x}_i) ) 
\end{aligned}
\end{equation}
Since the estimate is differentiable, it can be maximized via stochastic gradient descent \cite{Loaiza_2016}. Thus, our approach is to use the Generative Phase Space Reconstruction (GPSR) method \cite{Roussel_2023} with a normalizing flow instead of a conventional neural network. We call this approach MENT-Flow, in reference to the MENT algorithm.

It is not immediately obvious whether normalizing flows can learn complex 6D distributions from projections in reasonable time. Flows preserve the topological features of the base distribution; for example, flows cannot perfectly represent disconnected modes if the base distribution has a single mode \cite{Stimper_2022}. Thus, building complex flows requires layering transformations, either as a series of maps (discrete flows) or a system of differential equations (continuous flows), often leading to large models and expensive training.\footnote{A relevant example comes from Green, Ting, and Kamdar \cite{Green_2023}, who used continuous flows for 6D phase space density estimation from measured stellar phase space coordinates. Training times ranged from hours to days on a GPU, depending on the distribution complexity, with approximately $10^4$ particles per batch.} 

We found that neural spline flows (NSF) \cite{Durkan_2019} provide a sufficient blend of speed and power. In this model, the unnormalizing transformation $\mathcal{F}$ has the following autoregressive form:
\begin{equation}
    x_i = \tau (z_i; c_i(z_1, \dots, z_{i -1})),
\end{equation}
where $\vect{x} = [x_1, \dots , x_n]^T$, $\vect{z} = [z_1, \dots , z_n]^T$, and $\tau$ is an invertible function parameterized by $c_i(z_1, \dots, z_{i -1})$. The transformation is invertible for any $c_i$, and since $c_i$ depends only on the first $i - 1$ dimensions, the transformation has a triangular Jacobian matrix whose determinant can be computed efficiently. In the NSF model, the 1D transformer ($\tau$) is a monotonic rational-quadratic spline \cite{Durkan_2019}. The spline is defined by the locations of $K$ different knots and the derivative at each knot. These parameters are provided by the conditioner ($c$), a masked neural network \cite{Papamakarios_2017} in which connections between nodes in a regular feedforward neural network are removed to produce the triangular Jacobian matrix.

The model's representational power increases with the number of parameters in the masked neural network and the number of knots in the rational-quadratic splines. We can also define more than one flow layer. For the composition of $T$ layers
\begin{equation}
    \mathcal{F} = \mathcal{F}_T \circ \mathcal{F}_{T-1} \circ \dots \circ \mathcal{F}_{2} \circ \mathcal{F}_1,
\end{equation}
and transformed coordinates
\begin{equation}
    \vect{z}_t = \mathcal{F}_{t} (\vect{z}_{t - 1}),
\end{equation}
the Jacobian determinant is available from
\begin{equation}
    \left| \det J_{\mathcal{F}}(\vect{z}_0) \right| =
    \prod_{t=1}^T {
        \left| \det J_{\mathcal{F}_t}(\vect{z}_{t - 1}) \right|.
    }
\end{equation}

Compared to MENT, MENT-Flow increases the reconstruction model complexity and does not guarantee an exact entropy maximum. However, MENT-Flow scales straightforwardly to $n$-dimensional phase space and immediately generates independent and identically distributed samples from the reconstructed distribution function.

\section{Numerical experiments} \label{sec:experiments}

The following numerical experiments demonstrate that MENT-Flow solutions approach MENT solutions in 2D phase space. Subsequent experiments demonstrate that MENT-Flow can fit complicated 6D phase space distributions to large measurement sets in reasonable time and that entropic regularization keeps the reconstruction close to the prior. To simplify the examples, we focused on linear phase space transformations rather than more realistic accelerator models. We also tended to use ground-truth distributions without linear interplane correlations, highlighting nonlinear features.\footnote{Equivalently, we assume we know the covariance matrix $\vect{\Sigma} = \langle \vect{x}\vect{x}^T \rangle = \vect{V}\vect{V}^T$, where $\mathbf{V}$ is a symplectic matrix, and reconstruct the distribution in normalized coordinates $\vect{x}_n = \vect{V}^{-1}\vect{x}$ by setting $\mathcal{M}_k \rightarrow \mathcal{M}_k \vect{V}$. The covariance matrix is usually overdetermined by the measurements---for example, three measurements determine the $2 \times 2$ covariance matrix---so that all distributions that fit the data share the same covariance matrix. In these cases, it is reasonable to fit the covariance matrix first.} We chose to maximize the entropy relative to a Gaussian prior.\footnote{A Gaussian prior may be a reasonable choice for accelerator applications: (i) the prior has no interplane dependence and can expand to approximate a uniform distribution; (ii) any known elements of the $n \times n$ covariance matrix can be used to define the Gaussian prior; (iii) beams are typically clustered in phase space and approximately Gaussian at equilibrium; (iv) a Gaussian prior can be used to limit the beam size in dimensions that are weakly constrained by the data.} The flow's base distribution is also a Gaussian, so the entropy penalty pushes the flow toward an identity or scaling transformation.

Our normalizing flow architecture is described in the previous section. The flow consists of five layers. Each layer is an autoregressive transformation, where the 1D transformation along each dimension is a rational-quadratic spline with 20 knots; the function values and derivatives at the knots are parameterized by a masked neural network with 3 hidden layers of 64 hidden units. Note that increasing the model size should \textit{not} lead to overfitting since we train via maximum entropy, not maximum likelihood. 

We compare MENT-Flow to MENT. Our MENT implementation uses linear interpolation to evaluate the Lagrange functions at any location on the projection axes, and we simulate the projections by numerical integration. We also compare to an unregularized neural network (NN) whose only aim is to fit the data. The NN is a standard fully connected feedforward network with 3 hidden layers of 32 hidden units and $\tanh$ activation functions. 

We used $2 \times 10^4$ samples to estimate the entropy and projections. During each epoch, we trained the normalizing flow using 400 iterations of the Adam optimizer. After each epoch, we multiplied the penalty parameter by $1.5$. We stopped training when
\begin{equation}\label{eq:stop}
    \langle D \rangle = 
    \frac{1}{K}
    \sum_{k=1}^{K} {
        D[
            {g}_k(\vect{u}_{k_\parallel}), 
            \tilde{g}_k(\vect{u}_{k_\parallel})
        ]
    }
    < \epsilon,
\end{equation}
where $\langle D \rangle$ is the average divergence between the simulated and measured projections, $K$ is the number of measurements, and $\epsilon \approx 10^{-4}$ is a threshold. Other hyperparameter values are found in \cite{Hoover_2024_zenodo}, which contains the code to reproduce the figures in this paper.

\subsection{2D reconstructions from 1D projections}

Our first experiment tests the model performance in 2:1 phase space tomography. We assume an accelerator composed of drifts and quadrupole magnets, such that a symplectic transfer matrix $\vect{M}$ approximates the dynamics. The transfer matrix can be decomposed as
\begin{equation}\label{eq:floquet}
    \vect{M} = \vect{V}(\alpha_2, \beta_2) \vect{R}(\mu) \vect{V}(\alpha_1, \beta_1)^{-1},
\end{equation}
where 
\begin{equation}\label{eq:vmat}
    \vect{V}(\alpha, \beta) = 
    \begin{bmatrix}
        \sqrt{\beta} & 0 \\ -\frac{\alpha}{\sqrt{\beta}} & \frac{1}{\sqrt{\beta}}
    \end{bmatrix}
\end{equation}
is a normalization matrix, parameterized by $\alpha$ and $\beta$, and  
\begin{equation}\label{eq:rmat}
    \vect{R}(\mu) = \begin{bmatrix}
        \cos\mu & \sin\mu \\   \sin\mu & \cos\mu 
    \end{bmatrix}
\end{equation}
is a rotation by the phase advance $\mu$. The projection angle, and hence the reconstruction quality, depends only on the phase advance. Various constraints can limit the projection angle range, but we assume the projection angles are evenly spaced over the maximum 180-degree range. 

\begin{figure*}
    \centering
    \includegraphics[width=\textwidth]{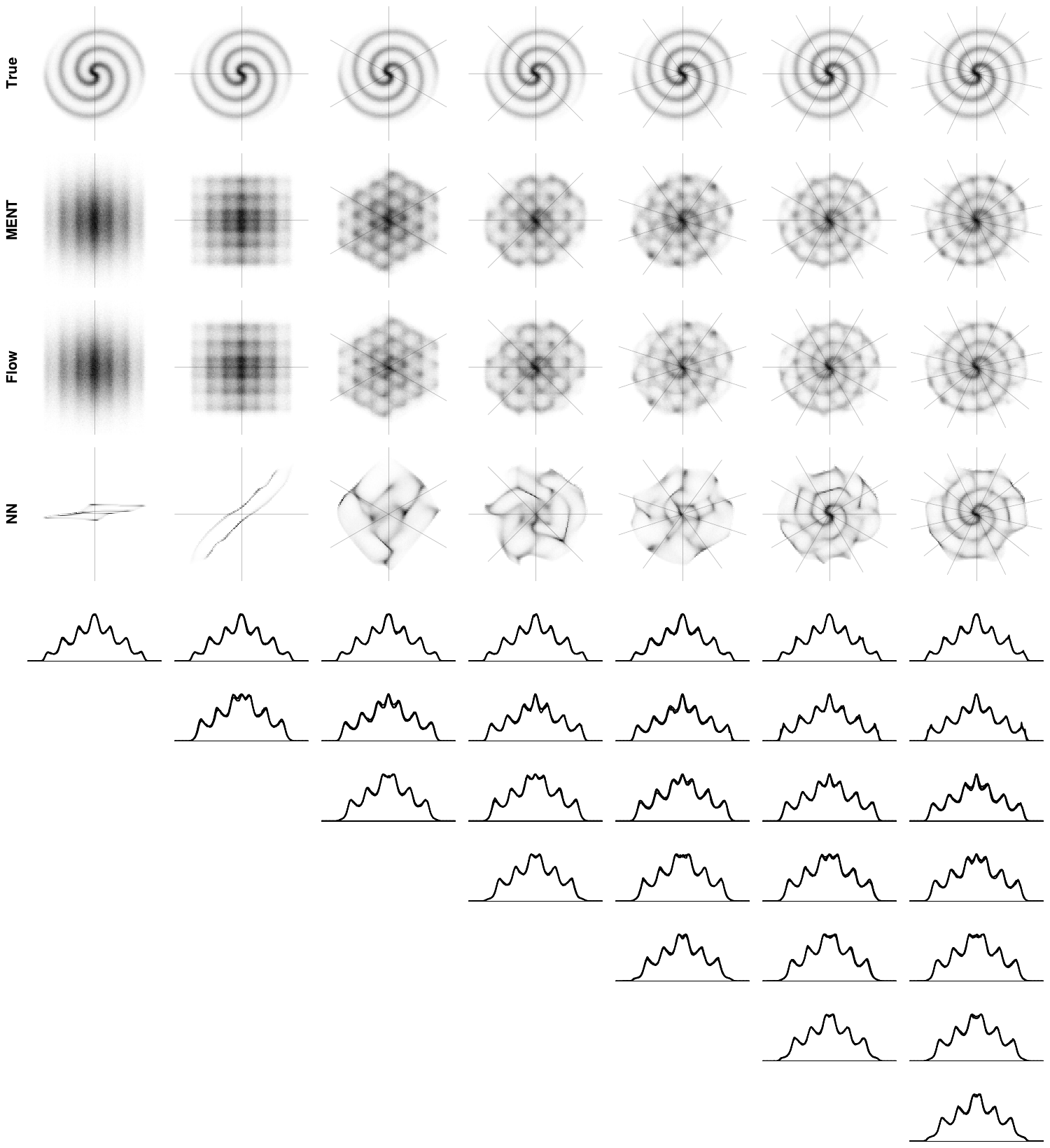}
    \caption{2D reconstructions from evenly spaced 1D projections. The top four rows plot samples from the true distribution, MENT reconstruction, MENT-Flow reconstruction, and NN reconstruction. Faint lines show the evenly spaced projection angles, increasing from 1 in the left column to 7 in the right column. In the bottom rows, the distributions are projected onto the measurement axes. (The four profiles overlap in most cases.)}
    \label{fig:rec_2d_two_spirals}
\end{figure*}

Fig.~\ref{fig:rec_2d_two_spirals} shows reconstructions from a varying number of projections, comparing MENT, MENT-Flow, and the unregularized neural network (NN).
It is clear that maximizing the stochastic estimate in Eq.~\eqref{eq:flow_entropy} pushes the distribution's entropy close to its constrained maximum. (Recall that MENT maximizes entropy by construction). Although the MENT solutions are of higher quality, the differences are not visible from afar.

Fig.~\ref{fig:rec_2d_two_spirals} illustrates that entropy maximization is a conservative approach to the reconstruction problem. All reconstructed features are implied by the data. In contrast, the distributions in the bottom rows fit the data but are unnecessarily complex. Of course, reconstructions from one or two projections are bound to fail if the prior is uninformative, but these cases are still useful because they demonstrate MENT's logical consistency: given only the marginal distributions and an uncorrelated prior, the posterior is the product of the marginals. On the other extreme, with enough data, the feasible distributions differ only in minor details. MENT shines in intermediate cases where the measurements contain just enough information to constrain the distribution's primary features. For example, the continuous spiral structure develops rapidly with the number of views in Fig.~\ref{fig:rec_2d_two_spirals}.

Fig.~\ref{fig:rec_2d_two_spirals} also illustrates the flow's capacity to represent complicated distributions despite the restriction to invertible transformations. This example focuses on spiral patterns, which are characteristic of nonlinear dynamics. (Additional examples are included in the supplemental material.) It is important to note that, while our analysis focuses on the beam core, low-density regions can also impact accelerator performance \cite{Aleksandrov_2021}. Flows can struggle to model distribution tails \cite{Laszkiewicz_2022_icml}. Our ground-truth distribution does not have significant halo and we do not report the agreement at this level; however, preliminary studies indicate the Kullback-Leibler (KL) divergence may enhance dynamic range relative to, i.e., the mean absolute error when fitting data.

\begin{figure}
    \centering
    \includegraphics[width=0.9\columnwidth]{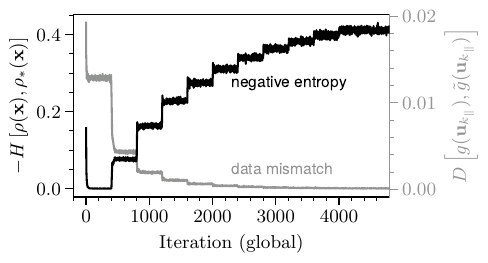}
    \caption{Entropy and data mismatch during training.}
    \label{fig:loss}
\end{figure}

Fig.~\ref{fig:loss} plots the entropy and data mismatch terms during training.
The end of each epoch is clear from the sharp jumps in the loss curves when the penalty parameter $\mu$ increases.
We aimed to keep the penalty parameter updates as small as possible. More aggressive update schedules did not lead to dramatically different results, but we did not explore this in detail. We did not see significant improvements using more sophisticated Augmented Lagrangian (AL) methods \cite{Loaiza_2016, Basir_2023}. A more important choice seems to be the stopping criteria, as increasing $\mu$ can eventually cause ill-conditioning. Our stopping condition ($\langle D \rangle \le 10^{-4}$ in Eq.~\eqref{eq:stop}) was chosen based on visual comparison of the simulated and true projections. We are not sure if the ideal stopping condition can be determined automatically.

\subsection{6D reconstructions from 1D projections}

It is more difficult to design and evaluate high-dimensional numerical experiments. \textit{First}, establishing reconstruction accuracy requires high-dimensional visualization or statistical distance metrics. We selected ground-truth distributions with clear high-dimensional structure and leveraged complete sets of pairwise projections and limited sets of partial projections (projections of slices) to aid the visualization.

\textit{Second}, we cannot determine the distance from the reconstructed distribution to the true maximum-entropy distribution without an analytic solution. We point to Fig.~\ref{fig:rec_2d_two_spirals} as evidence that the entropy penalty can push the MENT-Flow solution close to the exact solution. We also continued to train an unregularized neural network on the same data to show that additional solutions can exist far from the prior.

\textit{Third}, for a given beamline and a fixed number of measurements, we do not yet know how to find the information-maximizing set of 6D phase space transformations. In 2:1 tomography, if the transformations are linear, the reconstruction quality is tied to a single parameter (the projection angle). There is no such connection in $n$:2 tomography when $n > 3$, as there is no obvious analog of the projection angle in these cases. Here, to demonstrate the method, we instead restrict our attention to 1D projections. A 1D projection axis can be specified by a point on the unit sphere; if the distribution is spherically symmetric, we hypothesize that the optimal projection axes are uniformly spaced on the sphere. In 2D, this leads to evenly spaced projection angles between $0$ and $\pi$ radians. In our numerical experiments, we approximated this condition by randomly sampling points from a uniform distribution on the sphere. The points will \textit{not} be uniformly spaced, but in the limit of many projections, the reconstruction should converge to the true distribution \cite{Dai_2021_sinf}.

\begin{figure*}
    \centering
    \subfloat[][]{%
        \includegraphics[width=\columnwidth]{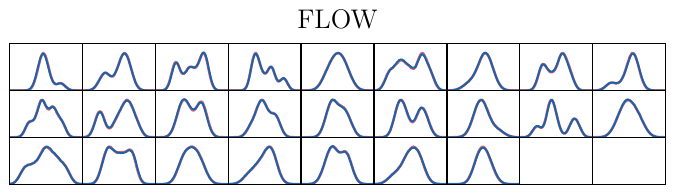}%
        \label{fig:rec_6d_1d_gmm_25meas-a}%
    }%
    \hfill
    \subfloat[][]{%
        \includegraphics[width=\columnwidth]{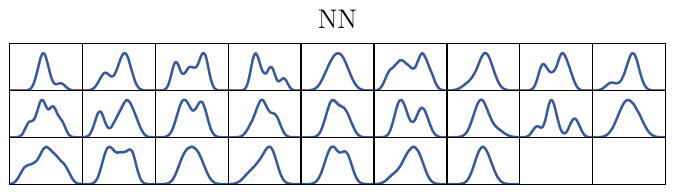}%
        \label{fig:rec_6d_1d_gmm_25meas-b}%
    }%
    \vfill
    \subfloat[][]{%
        \includegraphics[width=\columnwidth]{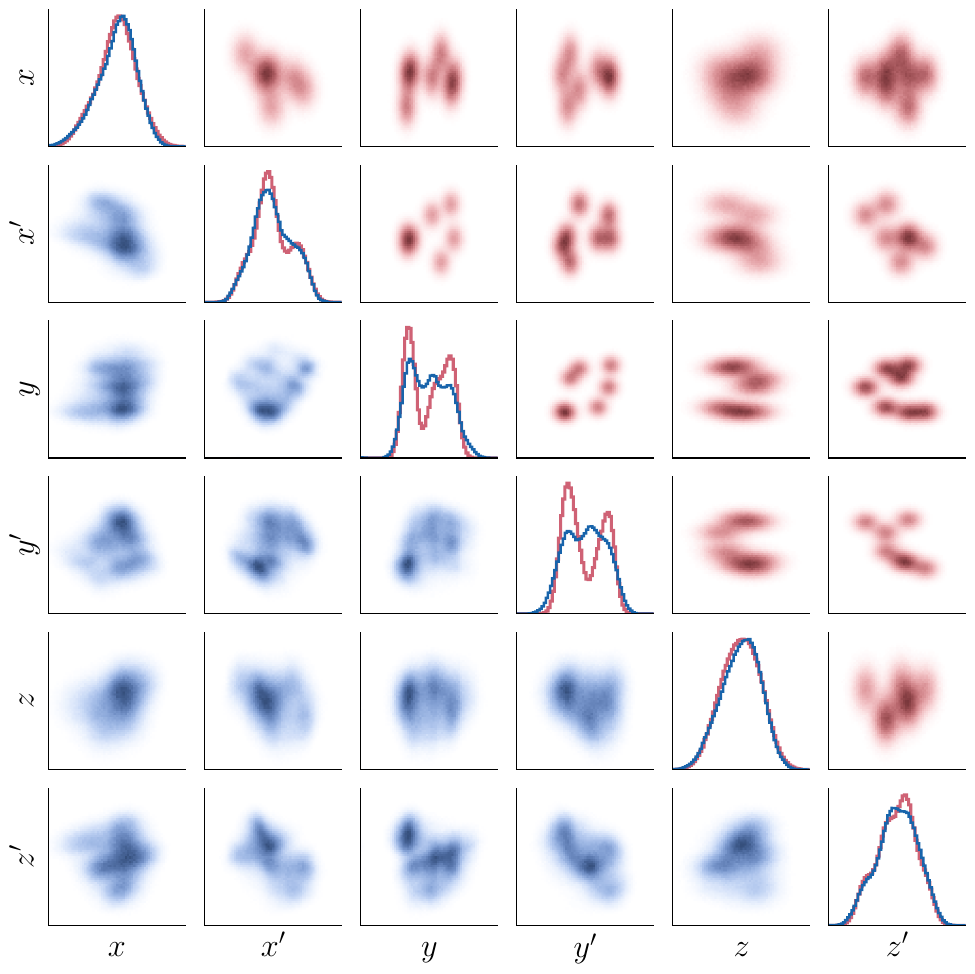}%
        \label{fig:rec_6d_1d_gmm_25meas-c}%
    }%
    \hfill
    \subfloat[][]{%
        \includegraphics[width=\columnwidth]{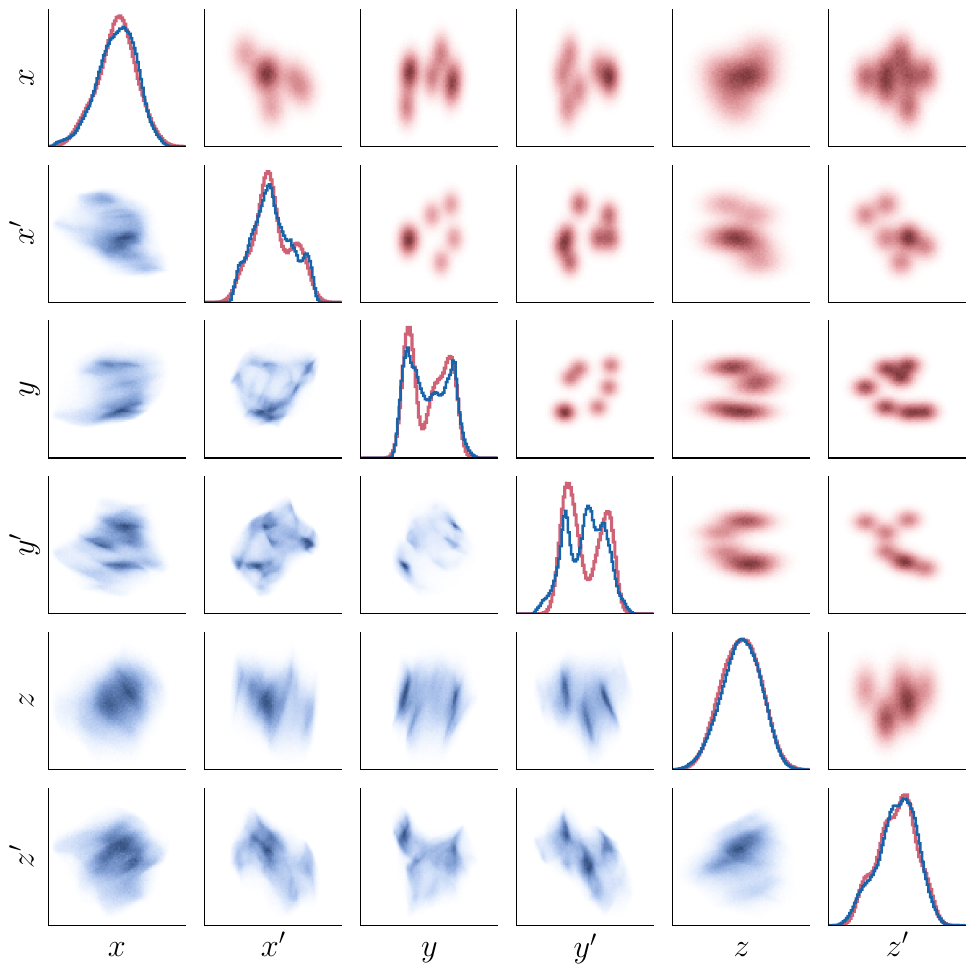}%
        \label{fig:rec_6d_1d_gmm_25meas-d}%
    }%
    \caption{Reconstruction of a 6D Gaussian mixture distribution from 25 random 1D projections. The MENT-Flow reconstruction on the left is compared to the NN reconstruction on the right. (a-b) Simulated projections (blue) vs. measured projections (red). (c-d) Low-dimensional views of the reconstructed distribution (blue) and the ground-truth distribution (red). 1D profiles are plotted on the diagonal subplots. 2D projections are plotted on the off-diagonal subplots.}
    \label{fig:rec_6d_1d_gmm_25meas}
\end{figure*}

\begin{figure*}
    \centering
    \subfloat[][]{%
        \includegraphics[width=\columnwidth]{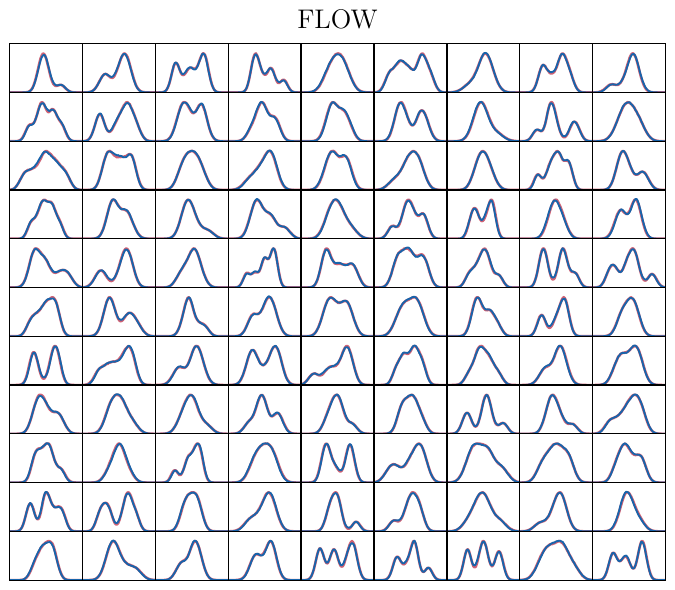}%
        \label{fig:rec_6d_1d_gmm_100meas-a}%
    }%
    \hfill
    \subfloat[][]{%
        \includegraphics[width=\columnwidth]{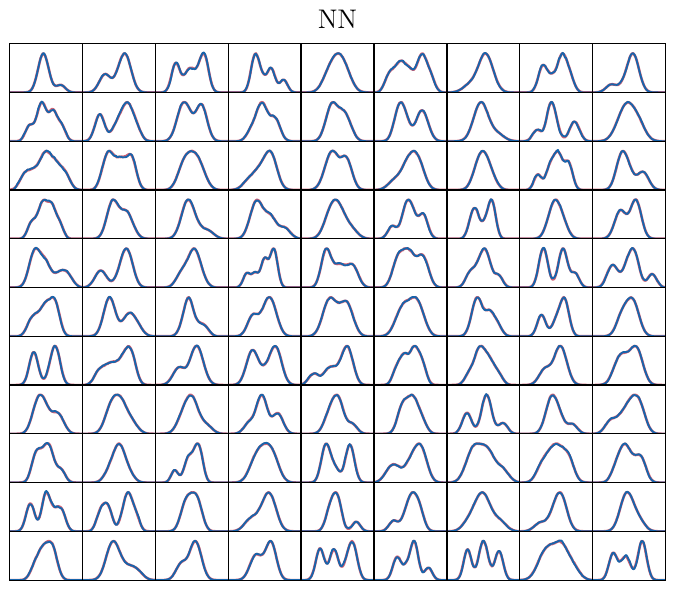}%
        \label{fig:rec_6d_1d_gmm_100meas-b}%
    }%
    \vfill
    \subfloat[][]{%
        \includegraphics[width=\columnwidth]{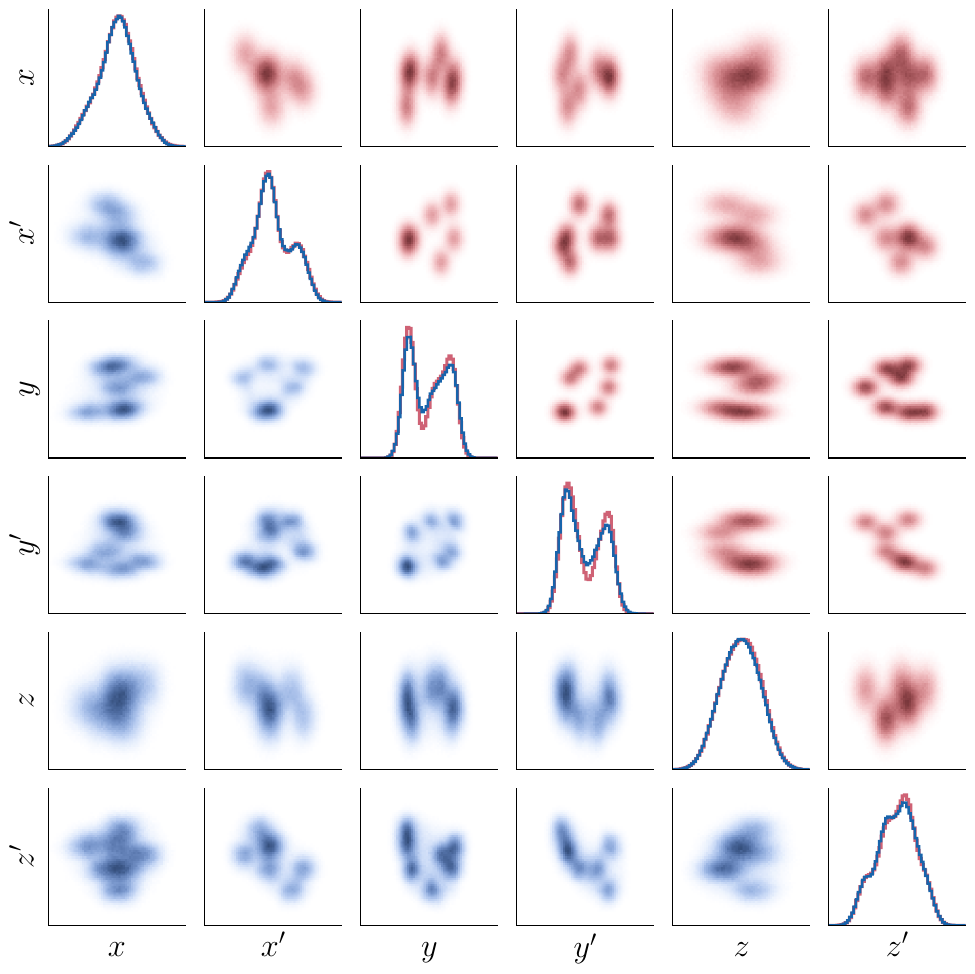}%
        \label{fig:rec_6d_1d_gmm_100meas-c}%
    }%
    \hfill
    \subfloat[][]{%
        \includegraphics[width=\columnwidth]{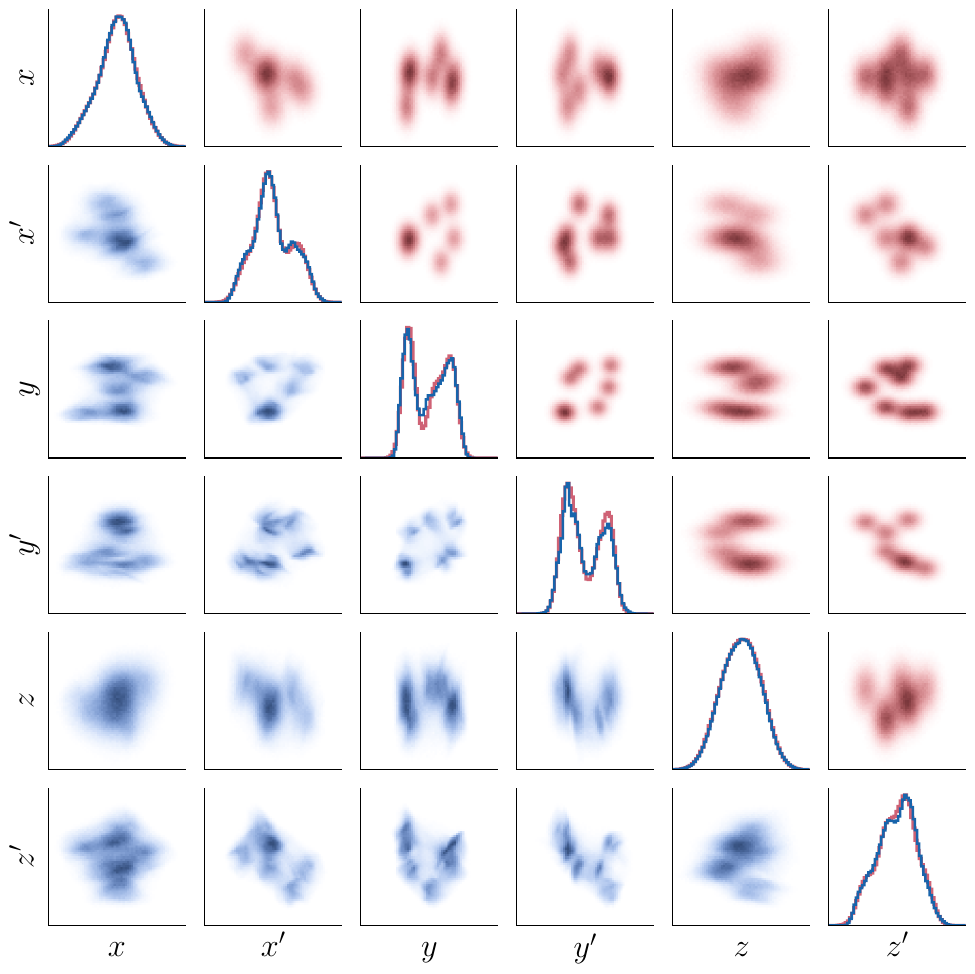}%
        \label{fig:rec_6d_1d_gmm_100meas-d}%
    }%
    \caption{Reconstruction of a 6D Gaussian mixture distribution from 100 random 1D projections. The MENT-Flow reconstruction on the left is compared to the NN reconstruction on the right. (a-b) Simulated projections (blue) vs. measured projections (red). (c-d) Low-dimensional views of the reconstructed distribution (blue) and the ground-truth distribution (red). 1D profiles are plotted on the diagonal subplots. 2D projections are plotted on the off-diagonal subplots.}
    \label{fig:rec_6d_1d_gmm_100meas}
\end{figure*}

Our first high-dimensional experiment, shown in Figs.~\ref{fig:rec_6d_1d_gmm_25meas}-\ref{fig:rec_6d_1d_gmm_100meas}, reconstructs a seven-mode Gaussian mixture distribution (a superposition of seven Gaussian distributions, each with a random mean and variance) from random 1D projections. Fig.~\ref{fig:rec_6d_1d_gmm_25meas} uses 25 projections and Fig.~\ref{fig:rec_6d_1d_gmm_100meas} uses 100 projections. This reconstruction used the same flow architecture as the 2D experiments. The NN architecture was changed to 2 layers of 50 units, still with $\tanh$ activation functions.
We draw the following conclusions. (i) Normalizing flows can represent complicated 6D distributions far from the unimodal base distribution. All simulated measurements match the training data. Charged particle beams are often smooth and unimodal, so this example represents a challenging case. Therefore, flow-based models are likely sufficient for many applications in accelerator physics. (ii) MENT-Flow can simultaneously fit a large number of measurements. (iii) The entropy penalty works as intended. The entropy-regularized solution fits the data just as well as the NN solution but eliminates high-frequency terms in the distribution function. The MENT-Flow solution is much closer to the smooth prior.

\begin{figure*}
    \centering
    \subfloat[][]{%
        \includegraphics[width=\columnwidth]{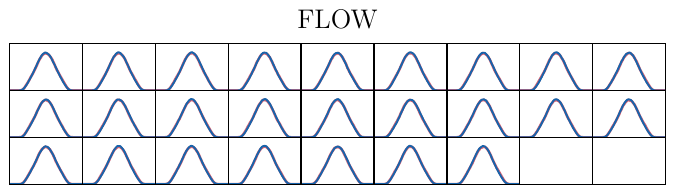}%
        \label{fig:rec_6d_1d_rings_25meas-a}%
    }%
    \hfill
    \subfloat[][]{%
        \includegraphics[width=\columnwidth]{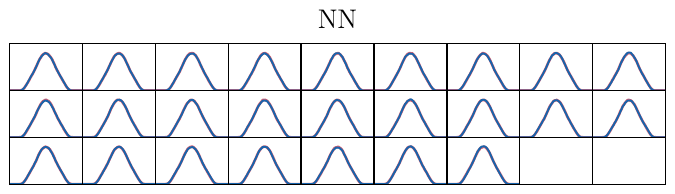}%
        \label{fig:rec_6d_1d_rings_25meas-b}%
    }%
    \vfill
    \subfloat[][]{%
        \includegraphics[width=\columnwidth]{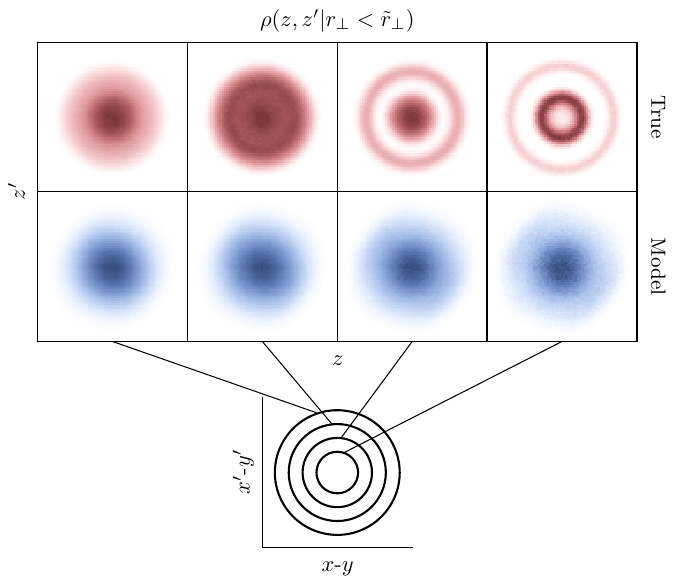}%
        \label{fig:rec_6d_1d_rings_25meas-e}%
    }%
    \hfill
    \subfloat[][]{%
        \includegraphics[width=\columnwidth]{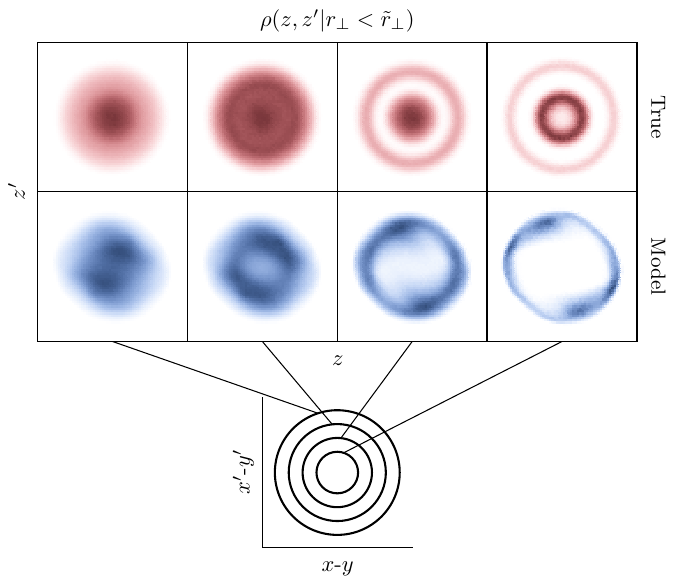}%
        \label{fig:rec_6d_1d_rings_25meas-f}%
    }%
    \caption{Reconstruction of 6D ``rings'' distribution from 25 random 1D projections. The MENT-Flow reconstruction on the left is compared to the NN reconstruction on the right. (a-b) Simulated projections (blue) vs. measured projections (red). (c-d) 2D projections of the 6D distribution $\rho(x, x', y, y', z, z')$ within a shrinking 4D ball in the $x$-$x'$-$y$-$y'$ plane. We define a ball of radius $\tilde{r}_\perp$ by ${r}_\perp \le \tilde{r}_\perp$, where ${r}_\perp = \sqrt{x^2 + x'^2 + y^2 + y'^2}$. Therefore, we write the projected density as $\rho(z, z' | r_\perp \le \tilde{r}_\perp)$. The ball shrinks from left to right. The largest radius (on the left) selects nearly all particles, while the smallest radius (on the right) selects particles near the core.}
    \label{fig:rec_6d_1d_rings_25meas}
\end{figure*}

\begin{figure*}
    \centering
    \subfloat[][]{%
        \includegraphics[width=\columnwidth]{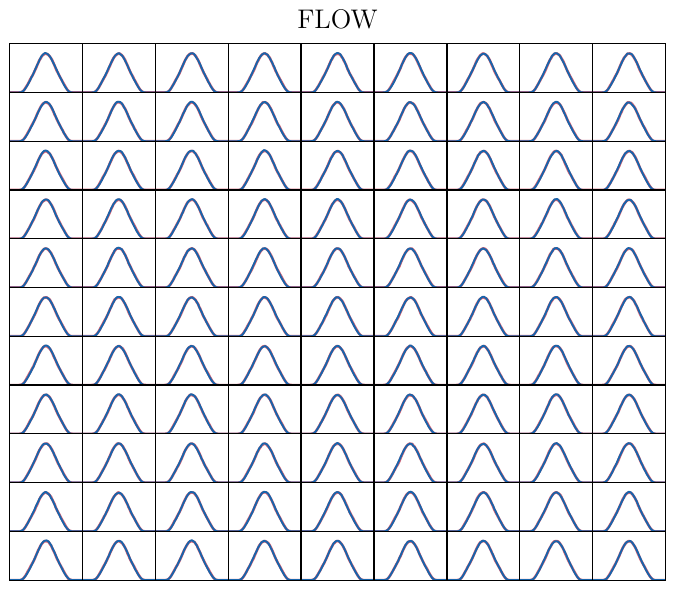}%
        \label{fig:rec_6d_1d_rings_100meas-a}%
    }%
    \hfill
    \subfloat[][]{%
        \includegraphics[width=\columnwidth]{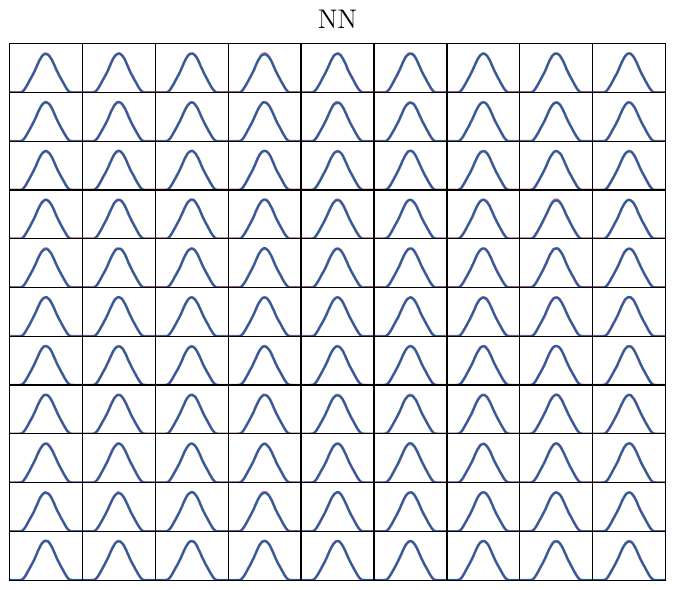}%
        \label{fig:rec_6d_1d_rings_100meas-b}%
    }%
    \vfill
    \subfloat[][]{%
        \includegraphics[width=\columnwidth]{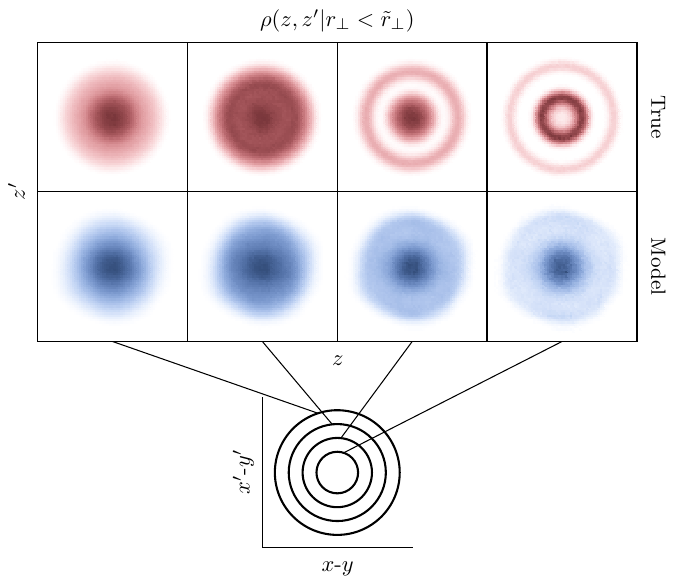}%
        \label{fig:rec_6d_1d_rings_100meas-e}%
    }%
    \hfill
    \subfloat[][]{%
        \includegraphics[width=\columnwidth]{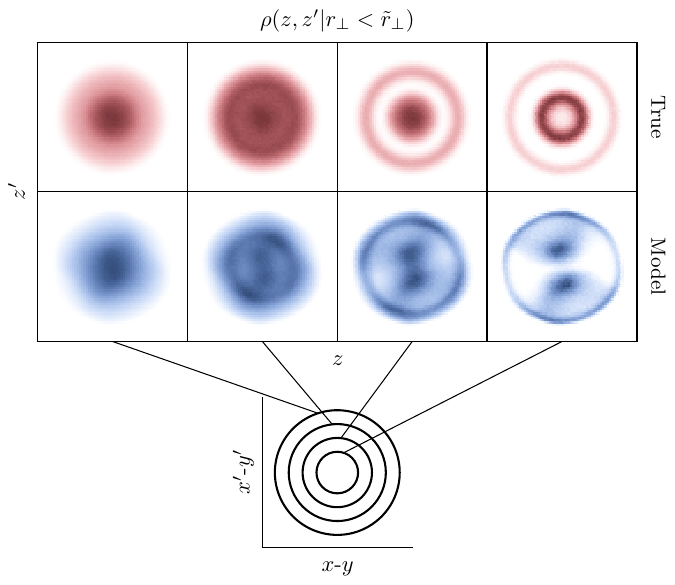}%
        \label{fig:rec_6d_1d_rings_100meas-f}%
    }%
    \caption{Reconstruction of 6D ``rings'' distribution from 100 random 1D projections. The MENT-Flow reconstruction on the left is compared to the NN reconstruction on the right. (a-b) Simulated projections (blue) vs. measured projections (red). (c-d) 2D projections of the 6D distribution $\rho(x, x', y, y', z, z')$ within a shrinking 4D ball in the $x$-$x'$-$y$-$y'$ plane. We define a ball of radius $\tilde{r}_\perp$ by ${r}_\perp \le \tilde{r}_\perp$, where ${r}_\perp = \sqrt{x^2 + x'^2 + y^2 + y'^2}$. Therefore, we write the projected density as $\rho(z, z' | r_\perp \le \tilde{r}_\perp)$. The ball shrinks from left to right. The largest radius (on the left) selects nearly all particles, while the smallest radius (on the right) selects particles near the core.}
    \label{fig:rec_6d_1d_rings_100meas}
\end{figure*}

The Gaussian mixture distribution has little overlap between modes, so mismatch between the true and reconstructed distribution is obvious from low-dimensional views. Hollow structures in high-dimensional phase space are not always evident from low-dimensional views. As an example, measurements at the Spallation Neutron Source (SNS) Beam Test Facility (BTF) show space-charge-driven hollowing in 3D and 5D projections of the 6D phase space distribution \cite{Cathey_2018, Ruisard_2020, Hoover_2023}. This motivates us to consider distributions with hidden internal structure. To this end, an $n$-dimensional ``rings'' distribution serves as the ground truth in Fig.~\ref{fig:rec_6d_1d_rings_25meas}-\ref{fig:rec_6d_1d_rings_100meas}; particles populate two concentric $n$-spheres with radii $r_2 = 2r_1$, and the radii are perturbed with Gaussian noise to generate a smooth density.

The entropy-regularized solution maintains the spherical symmetry of the Gaussian prior, flattening and eventually inverting its radial density profile to fit the data.\footnote{The one-dimensional projections in Figs.~\ref{fig:rec_6d_1d_rings_25meas}-\ref{fig:rec_6d_1d_rings_100meas} are nearly Gaussian. Klartag \cite{Klartag_2007} proved that almost all $m$-dimensional projections of an isotropic (no linear correlations) $n$-dimensional log-concave distribution function are nearly Gaussian when $n \gg m$. Many distributions commonly used in accelerator modeling are log-concave, such as the $n$-dimensional Gaussian, Waterbag (uniformly filled ball), and KV (uniformly filled sphere) distributions. A practical implication of this theorem is that small fluctuations in the $m$-dimensional projections have a greater impact on the $n$-dimensional reconstructed distribution as $n - m$ increases---for instance, completely inverting the density profile from peaked to hollow. Thus, we found that later training epochs can significantly change the distribution while only slightly decreasing the loss function. It follows that, for certain distributions, there may be some value of $n - m$ for which $n$:$m$ tomography is practically impossible.} The sliced views reveal an internal structure---a dense core surrounded by a low-density cloud---that MENT-Flow better approximates when measurements are scarce. In addition to injecting unnecessary correlations between planes, the unregularized solution ejects all particles from the core. Surprisingly, adding additional measurements does not solve the problem and generates two distinct modes in the reconstructed density. Using a different random seed to define the measurement axes can generate different patterns, but the hollowing and splitting just described are typical. Note that this internal structure is not obvious from the full 2D projections in the left column of Figs.~\ref{fig:rec_6d_1d_rings_25meas}-\ref{fig:rec_6d_1d_rings_100meas}.

\section{Conclusion and extensions} \label{sec:conclusion}

In conclusion, MENT-Flow is a promising approach to high-dimensional phase space tomography. Numerical experiments demonstrate consistency with known 2D maximum-entropy solutions and the ability to fit complex 6D distributions to large measurement sets. In the 6D tests, although there are no available benchmarks, we found that entropic regularization pulls the solution closer to the prior. Thus, MENT-Flow is an effective way to incorporate prior information in high-dimensional reconstructions. Our numerical experiments also emphasize the potential importance of uncertainty quantification in high-dimensional tomography, as we found that some distributions can only be reconstructed from large numbers of 1D measurements. Future work should apply MENT-Flow to more realistic distributions, accelerator models, and diagnostics, especially 2D projections. Future work should also aim to extend MENT to higher dimensions to serve as a benchmark.

MENT-Flow has several limitations. First, particle sampling is over 50\% slower than a conventional neural network, and the total runtime is inflated by the need to solve multiple subproblems to approach the maximum-entropy distribution from below. This motivates the search for more efficient flows and sample-based entropy estimates. Note that our training times ranged from 5 to 20 minutes on a single GPU, depending on the number of projections, phase space dimension, batch size, and penalty parameter updates. Second, MENT-Flow maximizes the entropy using a penalty method that does not generate exact solutions and requires a hand-tuned penalty parameter schedule to avoid ill-conditioning. It is unclear whether this process can be automated or whether alternative strategies can better prevent ill-conditioning. Third, MENT-Flow does not attach uncertainty to its output.

We now discuss possible extensions to new problems. First, it may be possible to fit $n$-dimensional distributions to $m$-dimensional projections when $m > 2$. This problem is of theoretical interest but also has some practical relevance. 3D and 4D projections can be measured relatively quickly using slit-screen-dipole measurement systems in low-energy hadron accelerators \cite{Cathey_2018, Ruisard_2020, Hoover_2023}. We propose to draw samples from the measured projections and minimize a differentiable statistical distance between these samples and samples from the normalizing flow.

Second, an interesting application of maximum-entropy tomography is to intense hadron beams, in which particles respond to both applied and self-generated electromagnetic fields. Phase space tomography is a significant challenge for such beams because the forward process depends on the unknown initial distribution. We do not know if the maximum-entropy distribution is unique in this case. Including space charge in the GPSR forward process may be possible using differentiable space charge solvers \cite{Qiang_2023}.

Finally, quantifying reconstruction uncertainty is a crucial step for the operational use of phase space tomography. In this paper, we defined a prior probability distribution $\rho(\vect{x})$ over the phase space coordinates $\vect{x}$. We may also define a prior over the space of distribution functions. In practice, the phase space distribution is parameterized by $\vect{\theta} \in \mathbb{R}^N$, so we may write the prior as $\mathcal{P}(\vect{\theta})$. The set of discretized measurements, i.e., histograms, can be expressed as a another parameter vector $\vect{d} \in \mathbb{R}^M$. Bayesian inference provides the update from prior to posterior:
\begin{equation}\label{eq:bayesian}
    \mathcal{P}(\vect{\theta} | \vect{d}) =
    \frac
    {
        \mathcal{P}(\vect{d} | \vect{\theta}) 
        \mathcal{P}(\vect{\theta}) 
    }
    {
        \mathcal{P}(\vect{d})
    },
\end{equation}
where $\mathcal{P}(\vect{d} | \vect{\theta})$ is the \textit{likelihood}, encoding the forward model, and $\mathcal{P}(\vect{d})$ is a normalizing constant. The principle of maximum entropy implies that we should prefer higher entropy phase space distributions. Gull and Skilling \cite{Skilling_1991} addressed this problem for image reconstruction and argued that the prior $\mathcal{P}(\vect{\theta})$ should be proportional to the exponential of the entropy $H(\vect{\theta})$. Given this prior, the strategy in this paper finds the maximum, or mode, of the posterior $\mathcal{P}(\vect{\theta} | \vect{d})$. The full posterior maps the entire solution space, encoding the reconstruction uncertainty. Future work could attempt to sample from a Gaussian approximation of the posterior at its maximum \cite{Skilling_1991}. Alternatively, it may be possible to sample from a more accurate approximation of the posterior using Markov Chain Monte Carlo (MCMC) or machine learning methods \cite{Mardani_2023}.

\section{Acknowledgements}

We are grateful to Ryan Roussel (SLAC National Accelerator Laboratory), Juan Pablo Gonzalez-Aguilera (University of Chicago), and Auralee Edelen (SLAC National Accelerator Laboratory) for discussions that seeded the idea for this work and for sharing their differentiable kernel density estimation code.

This manuscript has been authored by UT Battelle, LLC under Contract No. DE-AC05-00OR22725 with the U.S. Department of Energy. The United States Government retains and the publisher, by accepting the article for publication, acknowledges that the United States Government retains a non-exclusive, paid-up, irrevocable, world-wide license to publish or reproduce the published form of this manuscript, or allow others to do so, for United States Government purposes. The Department of Energy will provide public access to these results of federally sponsored research in accordance with the DOE Public Access Plan (http://energy.gov/downloads/doe-public-access-plan).

\bibliography{main}

\end{document}